\documentclass[12pt]{article}
\usepackage{amssymb}
\usepackage{amsmath}

\begin{document}

\title{ \begin{flushright}
{ \small CECS-PHY-07/14 }
\end{flushright}\vskip0.5cm Black holes, parallelizable horizons\\
and half-BPS states for the Einstein-Gauss-Bonnet\\
theory in five dimensions}
\author{Fabrizio Canfora$^{1,2}$, Alex Giacomini$^{1}$ and Ricardo Troncoso$%
^{1}$ \\
$^{1}$Centro de Estudios Cientificos (CECS), Casilla 1469 Valdivia, Chile.\\
$^{2}$Istituto Nazionale di Fisica Nucleare, Sezione di Napoli, GC Salerno.\\
e-mail: canfora@cecs.cl, giacomini@cecs.cl, ratron@cecs.cl}
\date{}
\maketitle

\begin{abstract}
Exact vacuum solutions with a nontrivial torsion for the
Einstein-Gauss-Bonnet theory in five dimensions are constructed. We consider
a class of static metrics whose spacelike section is a warped product of the
real line with a nontrivial base manifold endowed with a fully antisymmetric
torsion. It is shown requiring solutions of this sort to exist, fixes the
Gauss-Bonnet coupling such that the Lagrangian can be written as a
Chern-Simons form. The metric describes black holes with an arbitrary, but
fixed, base manifold. It is shown that requiring its ground state to possess
unbroken supersymmetries, fixes the base manifold to be locally a
parallelized three-sphere. The ground state turns out to be half-BPS, which
could not be achieved in the absence of torsion in vacuum. The Killing
spinors are explicitly found.
\end{abstract}

\section{Introduction}

Nowadays it is widely accepted by the high energy physics community that the
search for a unified theory seems to require additional spacetime
dimensions. In five dimensions, if one follows the basic principles of
General Relativity, the most general theory of gravity leading to second
order field equations for the metric is described by the so called
Einstein-Gauss-Bonnet action \cite{Lovelock}%
\begin{equation}
I=\kappa \int d^{5}x\sqrt{g}\left( R-2\Lambda +\tilde{\alpha} \left(
R^{2}-4R_{\mu \nu }R^{\mu \nu }+R_{\alpha \beta \gamma \delta
}R^{\alpha \beta \gamma \delta }\right) \right) \ ,  \label{Itensor}
\end{equation}%
where $\kappa $ is related to the Newton constant, $\Lambda $ to the
cosmological term, and $\tilde{\alpha} $ is the Gauss-Bonnet
coupling. For later convenience, it is useful to express the action
(\ref{Itensor}) in terms of
differential forms as \footnote{%
The relationship between the constants appearing in Eqs (\ref{Itensor}) and (%
\ref{action}) is given by $\tilde{\alpha} =\frac{c_{2}}{2c_{1}}$, $\Lambda =-6\frac{%
c_{0}}{c_{1}}$, $\kappa =2c_{1}$ . Moreover, for notational
simplicity, the wedge product between forms is understood.}

\begin{equation}
I=\int \epsilon _{abcde}\left( \frac{c_{0}}{5}e^{a}e^{b}e^{c}e^{d}e^{e}+%
\frac{c_{1}}{3}R^{ab}e^{c}e^{d}e^{e}+c_{2}R^{ab}R^{cd}e^{e}\right) \ ,
\label{action}
\end{equation}%
where $e^{a}=e_{\mu }^{a}dx^{\mu }$ is the vielbein, and $R^{ab}=d\omega
^{ab}+\omega _{\text{\ }c}^{a}\omega ^{cb}$ is the curvature $2$-form for
the spin connection $\omega ^{ab}=\omega _{\ \mu }^{ab}dx^{\mu }$.

In the first order formalism, the action (\ref{action}) is extremized
varying both with respect to the vielbein and with respect to the spin
connection independently, so that the field equations read%
\begin{equation}
\mathcal{E}_{e}:=\left(
c_{o}e^{a}e^{b}e^{c}e^{d}+c_{1}R^{ab}e^{c}e^{d}+c_{2}R^{ab}R^{cd}\right)
\epsilon _{abcde}=0\ ,  \label{equationcurvature}
\end{equation}%
and
\begin{equation}
\mathcal{E}_{ab}:=T^{c}\left( c_{1}e^{d}e^{e}+2c_{2}R^{de}\right) \epsilon
_{abcde}=0\ ,  \label{equationtorsion}
\end{equation}%
respectively. The torsion is defined as the covariant derivative of the
vielbein, i.e., $T^{a}=De^{a}$.

In the vanishing torsion sector, the field equations (\ref{equationtorsion})
are trivially fulfilled, and Eq. (\ref{equationcurvature})\ reduces to the
standard one in the second order formalism. Nevertheless, one peculiar
feature of the Einstein-Gauss-Bonnet theory, which distinguish it from
standard General Relativity, is that the field equations (\ref%
{equationtorsion}) do not necessarily imply the vanishing of torsion \cite%
{TZ-CQG}, so that in the first order formalism, besides the graviton, there
are additional propagating degrees of freedom related to the torsion.
Generically, the field equations (\ref{equationtorsion}) impose very strong
constraints on the torsion; however, this situation can be softened for
certain values of the Gauss-Bonnet coupling. For instance, requiring the
theory to posses the maximum number of degrees of freedom, fixes the
Gauss-Bonnet coupling as \cite{TZ-CQG}%
\begin{equation}
c_{2}=\frac{c_{1}^{2}}{4c_{0}}\ ,  \label{choice}
\end{equation}%
so that the theory possesses a unique maximally symmetric vacuum \cite%
{BH-Scan}, and the Lagrangian can be written as a Chern-Simons form \cite%
{Chamseddine}. For the choice (\ref{choice}) some exact solutions with
torsion\ in vacuum have been found \cite{Ar06}. Further choices of the
Gauss-Bonnet coupling also allow the existence of torsion in vacuum, for
which the number of degrees of freedom is not the maximum. Explicit
solutions of this sort have been recently found in \cite{CGW07}, which have
the structure of a cross product of a two-dimensional Riemannian manifold of
Lorentzian signature and constant curvature, with an Euclidean
tree-dimensional manifold of constant curvature and nonvanishing fully
antisymmetric torsion. Thus, it is natural to wonder whether there exists a
black hole solution of the five-dimensional Einstein-Gauss-Bonnet theory
whose horizon geometry is described by a three-dimensional manifold endowed
with a fully antisymmetric torsion.

In this paper, exact vacuum solutions with a nontrivial torsion for the
Einstein-Gauss-Bonnet theory in five dimensions are constructed. We consider
a class of static metrics of the form%
\begin{equation}
ds^{2}=-f^{2}\left( r\right) dt^{2}+\frac{dr^{2}}{f^{2}\left( r\right) }%
+r^{2}d\Sigma _{3}^{2}\ ,  \label{Ansatz-metric}
\end{equation}%
where $d\Sigma _{3}^{2}$ stands for the metric of a three-dimensional base
manifold $\Sigma _{3}$, which is endowed with a fully antisymmetric torsion.
In the next Section it is shown that requiring solutions of this sort to
exist, fixes the Gauss-Bonnet coupling as in Eq. (\ref{choice}), and the
metric describes black holes with an arbitrary, but fixed, base manifold $%
\Sigma _{3}$. In Section \ref{BPS} we show that requiring the black hole to
have a ground state possessing unbroken supersymmetries, fixes the base
manifold to be locally a parallelized three-sphere. It is worth pointing out
that unlike the Riemannian case, which is devoid of torsion, this ground
state turns out to be half-BPS, and their Killing spinors are explicitly
found. Finally, Section \ref{Discussion} is devoted to the discussion and
comments.

\section{Black holes with nontrivial torsion}

Let us now search for an exact solution of the field equations (\ref%
{equationcurvature}) and (\ref{equationtorsion}), whose metric is of the
form (\ref{Ansatz-metric}). The vielbein can then be chosen as
\begin{equation}
e^{0}=f(r)dt;\;\;\;e^{1}=f^{-1}(r)dr;\;\;\;e^{m}=r\tilde{e}^{m}\ ,
\label{ansatz}
\end{equation}%
where $\tilde{e}^{m}$ stands for the vielbein of the base manifold, so that
latin indices $m$, $n$, $p$, ... run along $\Sigma _{3}$.

The torsion is assumed to be fully antisymmetric, static, and such that its
only nonvanishing components have support only along the base manifold. A
simple ansatz in this case is given by%
\begin{equation}
T^{m}=K(r)\epsilon ^{mnp}e_{m}e_{p}\ .  \label{torsion}
\end{equation}%
It is useful to define the contorsion one-form $K^{ab}$, which fulfils $%
T^{a}=K^{ab}e_{b}$, so that the spin connection splits into its Riemannian
and non Riemannian parts according to
\begin{equation}
\omega ^{ab}=\mathring{\omega}^{ab}+K^{ab}\ ,  \label{spin-connection}
\end{equation}%
where $\mathring{\omega}^{ab}$ is the Levi-Civita (torsion-free) spin
connection, and the curvature two-form can be written as
\begin{equation}
R^{ab}=\mathring{R}^{ab}+\mathring{D}K^{ab}+K_{\ c}^{a}K^{cb}
\label{Curvature+DK+KK}
\end{equation}%
where $\mathring{R}^{ab}=d\mathring{\omega}^{ab}+\mathring{\omega}_{\text{\ }%
c}^{a}\mathring{\omega}^{cb}$ is the Riemannian curvature, and $\mathring{D}$
is the covariant derivative for the Levi-Civita spin connection.

Hence, by virtue of Eq. (\ref{torsion}), the only non vanishing components
of the contorsion are given by%
\begin{equation}
K^{mn}=-K(r)\epsilon ^{mnp}e_{p}\ ,  \label{contorsion0}
\end{equation}%
so that the spin connection for the ansatz (\ref{ansatz}), and (\ref{torsion}%
) reads

\begin{equation}
\omega ^{01}=f^{\prime }e^{0};\;\;\;\omega ^{n1}=f(r)\tilde{e}%
^{n};\;\;\;\omega ^{mn}=\tilde{\omega}^{mn}+K^{mn},  \label{levicivita}
\end{equation}%
where $\tilde{\omega}^{mn}$ corresponds to the Levi-Civita connection of the
base manifold.

The Riemannian curvature reads
\begin{gather}
\mathring{R}^{01}=-\frac{\left( f^{2}\right) ^{\prime \prime }}{2}%
e^{0}e^{1}\;\;\;;\;\;\;\mathring{R}^{0n}=-\frac{(f^{2})^{\prime }}{2r}%
e^{0}e^{n}  \notag \\
\mathring{R}^{1n}=-\frac{(f^{2})^{\prime }}{2r}e^{1}e^{n}\;\;\;;\;\;\;%
\mathring{R}^{mn}=\tilde{R}^{mn}-\frac{f^{2}}{r^{2}}e^{m}e^{n}
\label{riemann}
\end{gather}%
where $\tilde{R}^{mn}$ stands for the curvature of the base manifold $\Sigma
_{3}$, so that by virtue of (\ref{Curvature+DK+KK}), the curvature two-form
is given by%
\begin{gather}
R^{01}=\mathring{R}^{01}\;\;\;;\;\;\;R^{0m}=\mathring{R}^{0m}\;\;\;;\;\;%
\;R^{1n}=\mathring{R}^{1n}-\frac{f}{r}T^{n}\ ,  \notag \\
R^{mn}=\mathring{R}^{mn}-\frac{d\left( rK\right) }{r}\epsilon
^{mnp}e_{p}-K^{2}e^{m}e^{n}\ .  \label{curv+tors}
\end{gather}

\bigskip

Let us begin solving the field equations (\ref{equationtorsion}) assuming a
nonvanishing torsion.

\bigskip

Due to the form of our ansatz (\ref{Ansatz-metric}) and (\ref{torsion}), the
components $\mathcal{E}_{1m}=0$, and $\mathcal{E}_{0m}=0$ of the field
equations (\ref{equationtorsion}) are identically fulfilled. The components
of (\ref{equationtorsion}) along the base manifold, i.e., $\mathcal{E}%
_{mn}=0 $ are solved provided $\left( f^{2}\right) ^{\prime \prime
}=c_{1}/c_{2}$, which means that the function $f^{2}(r)$ is of the form%
\begin{equation}
f^{2}=\frac{c_{1}}{2c_{2}}r^{2}+\alpha r-\mu \ .  \label{lapse}
\end{equation}%
The remaining equation $\mathcal{E}_{01}=0$, is solved provided $d(rK)=0$,
which fixes the form of the function $K(r)$ in Eq. (\ref{torsion}) as
\begin{equation}
K=-\frac{\delta }{r}\ ,  \label{1torsi1}
\end{equation}%
where $\delta $ is an integration constant.

\bigskip

Let us now focus on the field equations (\ref{equationcurvature}).

\bigskip

The radial component of (\ref{equationcurvature}), $\mathcal{E}_{1}=0$,
reduces to%
\begin{equation}
6A(r)+\tilde{R}B(r)=0\ ,  \label{Eps1}
\end{equation}%
where $\tilde{R}$ is the Ricci scalar of the base manifold, and%
\begin{eqnarray*}
A &=&4c_{0}-\frac{c_{1}^{2}}{c_{2}}+2c_{2}\alpha \left( \frac{\delta
^{2}-\mu }{r^{3}}+\frac{\alpha }{r^{2}}\right) \ , \\
B &=&-2\frac{c_{2}\alpha }{r}\ .
\end{eqnarray*}%
Since $\tilde{R}$ depends only on the coordinates of the base manifold, Eq. (%
\ref{Eps1}) implies that%
\begin{equation}
A+\gamma B=0\ ,  \label{A=-gammaB}
\end{equation}%
where $\gamma $ is a constant.

\bigskip

Hence, Eq. (\ref{A=-gammaB}) implies that the constant $\alpha $ vanishes,
and that the Gauss-Bonnet coupling must necessarily be fixed as in Eq. (\ref%
{choice}). Furthermore, note that Eq. (\ref{Eps1}) does not impose any
restriction on the base manifold $\Sigma _{3}$. Using this, it is easy to
verify that the constraint $\mathcal{E}_{0}=0$, as well as the remaining
field equations $\mathcal{E}_{m}=0$ are fulfilled.

\bigskip

In sum, the solution describes an asymptotically AdS spacetime whose metric
reads%
\begin{equation}
ds^{2}=-\left( \frac{r^{2}}{l^{2}}-\mu \right) dt^{2}+\frac{dr^{2}}{\frac{%
r^{2}}{l^{2}}-\mu }+r^{2}d\Sigma _{3}^{2}\ ,  \label{Black hole}
\end{equation}%
where $l=\sqrt{\frac{2c_{2}}{c_{1}}}$ is the AdS radius\footnote{%
The asymptotically dS solution is obtained just making $l\rightarrow il$. In
this case the solution generically possesses a cosmological horizon and a
timelike naked (Riemannian) curvature singularity at the origin. The torsion
also diverges at the origin, even in the case $\mu =-1$, for which the
metric reduces to de Sitter spacetime. We no longer discuss this case here.}%
, possessing a non vanishing torsion whose only nonvanishing components are
given by%
\begin{equation}
T^{m}=-\frac{\delta }{r}\epsilon ^{mnp}e_{m}e_{p}\ .  \label{Torsion-BH}
\end{equation}

For this solution, the torsion and the Riemannian curvature are singular at
the origin, but nevertheless these singularities are surrounded by an event
horizon located at $r^{2}=\mu l^{2}$, so that this geometry describes a
black hole whose horizon geometry is endowed with a nontrivial torsion.

It is worth to remark that the field equations are solved for any fixed base
manifold $\Sigma _{3}$.

In the absence of torsion, this solution has been recently shown to exist
for the Einstein-Gauss-Bonnet theory only if the coefficients are fine tuned
as in Eq. (\ref{choice}) \cite{DOT2}. For base manifolds of constant
curvature the torsionless solution reproduces the ones found in \cite%
{Cai-Soh}, and \cite{ATZ}, which in the case of spherical symmetry reduces
to the one in \cite{BD}, \cite{BTZ}\footnote{%
For a generic value of the Gauss-Bonnet coupling $c_{2}$, the solution with
a base manifold of constant curvature in the absence of torsion has been
found in \cite{Cai}.}.

It is worth pointing out the similarity of the black hole solution found
here with the BTZ black hole \cite{Banados:1992wn}, \cite{Banados:1992gq}.
In three dimensions the static solution describes a black hole provided $\mu
>0$, and $\mu =0$ is the "black hole vacuum". For the range $-1<\mu <0$, the
solution has a naked conical singularity, and for $\mu =-1$ the metric is
that of $AdS_{3}$. For the static case, the only solutions with unbroken
supersymmetries within this family are $AdS_{3}$, being maximally
supersymmetric, and the zero mass black hole which has two Killing spinors
\cite{Coussaert-Henneaux}, so that it corresponds to the ground state of $%
(1,1)$-AdS supergravity \cite{AT} with periodic (Ramond) boundary conditions
on the spinor fields.

In the case of the five-dimensional black hole without torsion, the
situation is almost analogous to the three-dimensional case, since the
solution for $\mu =-1$ corresponds to $AdS_{5}$ spacetime, while for the
range $-1<\mu <0$, it describes timelike naked singularities. The black hole
is also obtained for $\mu >0$, and the solution with $\mu =0$ corresponds to
the black hole vacuum. However, it has been shown that in the absence of
torsion and matter fields, the only solution possessing Killing spinors
within this family is the maximally supersymmetric $AdS_{5}$ spacetime, so
that the zero mass black hole breaks all the supersymmetries \cite%
{SUSY-ground}.

Hence, it is natural to wonder whether the presence of torsion helps to
improve the situation in five dimensions, in the sense that if the black
hole vacuum with torsion had Killing spinors, its stability would be
guaranteed preventing the black hole from decaying into naked singularities.

In the next section it is shown that for the black hole solution given by
Eqs. (\ref{Black hole}) and (\ref{Torsion-BH}), which possesses an arbitrary
but fixed base manifold, requiring its groundstate to possess unbroken
supersymmetries, removes the arbitrariness in the base manifold since $%
\Sigma _{3}$ becomes fixed to be locally a parallelized (combed)
three-sphere. It is worth to remark that unlike the standard torsion-free
case, this ground state turns out to be half-BPS.

\section{Half-BPS ground state and locally parallelizable $S^{3}$ horizon}

\label{BPS}

The locally supersymmetric extension of the Einstein-Gauss-Bonnet theory in
five dimensions with the choice of couplings as in (\ref{choice}) is known
\cite{Chamseddine2}. The field equations as well as the local supersymmetry
transformations can be explicitly found in \cite{TZallOdd} (see also \cite%
{MsTZ}). It is simple to prove that our solution, given by Eqs. (\ref{Black
hole}) and (\ref{Torsion-BH}), solves the field equations of supergravity in
the absence of matter fields.

The Killing spinor equation is obtained requiring a purely bosonic
configuration to possess unbroken global supersymmetries. In the absence of
matter fields, i.e., for the purely gravitational sector, the Killing spinor
equation is given by%
\begin{equation}
\nabla \epsilon :=\left( d+\frac{1}{4}\omega ^{ab}\Gamma _{ab}+\frac{1}{2l}%
e^{a}\Gamma _{a}\right) \epsilon =0\ ,  \label{KS-Eq}
\end{equation}%
where $\Gamma_{ab}$ is expressed in terms of Dirac matrices as
$\Gamma _{ab}=(1/2)[ \Gamma _{a},\Gamma _{b}]$.
The consistency condition of Eq. (\ref{KS-Eq}) reads%
\begin{equation}
\nabla \nabla \epsilon =\left( \frac{1}{4}\left( R^{ab}+\frac{1}{l^{2}}%
e^{a}e^{b}\right) \Gamma _{ab}+\frac{1}{2l}T^{a}\Gamma _{a}\right) \epsilon
=0\ .  \label{Consistency-KS}
\end{equation}%
For the black hole solution given by Eqs. (\ref{Black hole}) and (\ref%
{Torsion-BH}), where the local frame has been chosen as in Eq. (\ref{ansatz}%
), the consistency condition turns out to be%
\begin{equation}
\left[ \frac{1}{2}\left( \tilde{R}^{mn}+(\mu -\delta ^{2})\tilde{e}^{m}%
\tilde{e}^{n}\right) \Gamma _{mn}+r\delta \epsilon ^{mnp}\tilde{e}_{n}\tilde{%
e}_{p}\left( \frac{1}{l}-\sqrt{\frac{1}{l^{2}}-\frac{\mu }{r^{2}}}\Gamma
_{1}\right) \Gamma _{m}\right] \epsilon =0\ .  \label{0coco0}
\end{equation}%
Note that there is no way to collect terms having the same functional
dependence on $r$ \textit{unless} $\mu =0$. In this case the consistency
condition reduces to%
\begin{equation}
\left[ \frac{1}{2}\left( \tilde{R}^{mn}-\delta ^{2}\tilde{e}^{m}\tilde{e}%
^{n}\right) \Gamma _{mn}+\frac{r}{l}\delta \epsilon ^{mnp}\tilde{e}_{n}%
\tilde{e}_{p}\Gamma _{m}\left( 1+\Gamma _{1}\right) \right] \epsilon =0\ ,
\label{Cond2}
\end{equation}%
which admits nontrivial solutions provided the Killing spinor is an
eigenstate of $\Gamma _{1}$, i.e.,%
\begin{equation}
\Gamma _{1}\epsilon =-\epsilon \ .  \label{Chirality}
\end{equation}%
Hence, Eq. (\ref{Cond2}) reduces to a condition on the Riemaniann curvature
of the base manifold given by%
\begin{equation}
\left( \tilde{R}^{mn}-\delta ^{2}\tilde{e}^{m}\tilde{e}^{n}\right) \Gamma
_{mn}\epsilon =0\ .  \label{Cond-sigma}
\end{equation}%
Since the base manifold $\Sigma _{3}$ is of Euclidean signature, this last
equation implies that $\Sigma _{3}$ must be locally a space of positive
constant curvature, whose radius is fixed in terms of the strength of the
torsion as%
\begin{equation}
\tilde{R}^{mn}=\delta ^{2}\tilde{e}^{m}\tilde{e}^{n}\ .
\label{Fixed Base manifold}
\end{equation}%
Therefore, as the base manifold describes the geometry of the horizon,
orientability and smoothness has to be required. Condition (\ref{Fixed Base
manifold}) then fixes the Riemannian geometry of $\Sigma _{3}$ to be that of
any smooth and orientable quotient of the three-sphere $S^{3}$, which are
fully classified (see e. g. \cite{Wolf}). A simple example corresponds to
the real projective space $RP^{3}$. Furthermore, as it can be seen from Eq. (%
\ref{curv+tors}), in this case the curvature two-form of the base manifold
vanishes, and hence, the base manifold is fixed to be locally a parallelized
three-sphere. Thus, the vielbeins of the base manifold can be chosen locally
as $\tilde{e}^{m}=\sigma ^{m}$, where $\sigma ^{m}$ stand for the
left-invariant forms of $SU(2)$, satisfying $d\sigma ^{m}=-$ $\epsilon
^{mnp}\sigma _{n}\sigma _{p}$, so that the Levi-Civita spin connection of $%
\Sigma _{3}$ reads $\tilde{\omega}^{mn}=-\epsilon ^{mnp}\tilde{e}_{p}$.

Note that for the generic black hole solution given by (\ref{Black hole})
and (\ref{Torsion-BH}) the constant $\delta $ is arbitrary. However, the
condition (\ref{Fixed Base manifold}) implies that $\delta $ cannot longer
be \textquotedblleft a truly" integration constant, since it can be brought
into the form $\delta =1$ under suitable rescalings of time and radial
coordinates\footnote{%
The case $\delta =-1$ is connected to the case $\delta =1$ trough a full
reflection of the local frames of the base manifold. It is then convenient
to choose $\delta =1$, so that the base manifold can be manifestly
paralellized making its intrinsic spin connection to vanish.}.

\bigskip

Let us solve the Killing spinor equation (\ref{KS-Eq}). The consistency
condition fixes $\mu =0$, so that $f=r/l$, and as the base manifold becomes
parallelized, its intrinsic spin connection vanishes (see Eq. (\ref%
{levicivita})). Furthermore, using the chirality condition (\ref{Chirality}%
), and reading the vielbein and the remaining components of the spin
connection from Eqs. (\ref{ansatz}), (\ref{spin-connection}), (\ref%
{Torsion-BH}) with $\delta =1$, and (\ref{levicivita}), respectively, the
Killing spinor equation reduces to

\begin{equation}
\left( d-\frac{1}{2r}dr\right) \epsilon =0\ .  \label{KS-ready}
\end{equation}%
Therefore, the Killing spinor does not depend neither on time nor on the
coordinates of the base manifold (since $\partial _{t}\epsilon =\partial
_{m}\epsilon =0$). The Killing spinors are given by the solution radial
equation%
\begin{equation}
\left( \partial _{r}-\frac{1}{2r}\right) \epsilon =0\ ,  \label{rcomp}
\end{equation}%
which can be integrated as%
\begin{equation}
\epsilon =\sqrt{\frac{r}{l}}\eta _{0}\ ,  \label{Killing spinors}
\end{equation}%
where $\eta _{0}$ is a constant spinor satisfying the chirality condition
\begin{equation}
\Gamma _{1}\eta _{0}=-\eta _{0}\ .  \label{Chiral-eta-zero}
\end{equation}

\bigskip

Therefore, unlike for the torsion-free case, one concludes that the black
holes described by Eqs. (\ref{Black hole}) and (\ref{Torsion-BH}), which
generically have an arbitrary but fixed base manifold, possess a half-BPS\
groundstate only for base manifolds which are locally a parallelized
(combed) three-sphere with a torsion given by%
\begin{equation}
T^{m}=-\frac{1}{r}\epsilon ^{mnp}e_{m}e_{p}\ .  \label{Torsion-parallel}
\end{equation}%
The supersymmetry of the groundstate, which would guarantee its stability,
then prevents a black hole with a torsion given by (\ref{Torsion-parallel})
from decaying into naked singularities. The Killing spinors are explicitly
given by Eq. (\ref{Killing spinors}) with the chirality condition (\ref%
{Chiral-eta-zero})\footnote{In view of the highly non-trivial effect
of the torsion, it would be interesting to compute its coupling with
the spin current as done in \cite{camella}}.

\section{Discussion and comments}

\label{Discussion}

In this paper we have found an exact black hole solution in vacuum with
nontrivial torsion for the Einstein-Gauss-Bonnet theory in five dimensions.
The metric is given by Eq. (\ref{Black hole}), and this spacetime is endowed
with torsion as in Eq. (\ref{Torsion-BH}). It was shown that requiring
solutions of this sort to exist, fixes the Gauss-Bonnet coupling as in Eq. (%
\ref{choice}), such that the Lagrangian can be written as a Chern-Simons
form. The metric describes black holes with an arbitrary, but fixed, base
manifold, and it is such that in the torsion-free the solution reduces to
the one found in \cite{DOT2}.

It is simple to prove that the black hole with torsion solves the
supergravity field equations in the absence of matter fields\footnote{%
Exact solutions of the supergravity field equations with a Gauss-Bonnet
coupling of the form (\ref{choice}), and with nonvanishing matter fields can
be found in Refs. \cite{ChTZ}, \cite{Banados}, \cite{MsTZ}, \cite{Toppan}.
\par
{}}. Thus, it was shown that requiring the black hole to have a groundstate
with unbroken supersymmetries, fixes the integration constant of the torsion
as in Eq. (\ref{Torsion-parallel}), so that the base manifold is constrained
to be locally a parallelized three-sphere. It is worth to remark that unlike
the standard torsion-free case, this ground state turns out to be half-BPS,
which would guarantee its stability, preventing a black hole with a torsion
given by (\ref{Torsion-parallel}) from decaying into naked singularities.
The Killing spinors are given explicitly by Eq. (\ref{Killing spinors}) with
the chirality condition (\ref{Chiral-eta-zero}).

It is worth to remark that the possibility of having a non trivial torsion
in vacuum may have a very important role in stabilizing other classes of
solutions which otherwise would be unstable.

Note that when the Gauss-Bonnet coupling is chosen as (\ref{choice}) the
theory not only acquires propagating degrees of freedom for the torsion, but
it is also invariant under a local $AdS$ boosts \cite{TZ-CQG}, so that the
curvature and the torsion transform into each other in a non trivial way.
Thus, a torsionless solution can be transformed into another one with
torsion given by%
\begin{equation*}
\delta T^{a}=\left( R^{ab}+\frac{1}{l^{2}}e^{a}e^{b}\right) \lambda _{b}\ ,
\end{equation*}%
where $\lambda _{b}$ is the $AdS$ boost parameter, which is not
diffeomorphically equivalent to the former.

In this sense the black hole solution found here possesses an intrinsic
torsion, since it cannot be gauged away with an $AdS$ boost.

It is worth pointing out that the ansatz for the torsion
(\ref{torsion}) has been inspired by an analogy between spacetimes
with nontrivial torsion and BPS\ states in Yang-Mills theory
\cite{Ca07}, which suggests the existence of further possibilities
to find exact solutions with intrinsic torsion. It would also be
interesting to see whether this kind of torsion contributes to the
spin current such that it could affect the chiral anomaly as in Ref.
\cite{camella}.

It is also natural to wonder whether there are black holes with a
non trivial torsion in vacuum for dimensions other than five. It has
been recently shown that three-dimensional supergravity can be
extended to the
case in which the gravitational sector admits torsion in vacuum \cite%
{Camelio}, so that the BTZ black hole with torsion \cite{ALBERTO} solves the
field equations in vacuum. Thus, the Killing spinors of the extremal cases
of the BTZ black hole with torsion can be obtained from the ones of the BTZ
black hole without torsion \cite{Coussaert-Henneaux}, by means of a simple
map introduced in \cite{Camelio}.

It would be interesting to explore the possibility of having unbroken
supersymmetries in dimensions higher than five with non trivial torsion in
vacuum, which is allowed for the class of supergravity theories constructed
in Refs. \cite{7-11}, and \cite{TZallOdd}.

\bigskip

\textit{Acknowledgments.-- }We thank J. Oliva for useful remarks. Special
acknowledge to Steve Willison, who participated in an early stage of this
work, for helpful comments. F.C. thanks PRIN SINTESI 2007 for financial
support. This work was partially funded by FONDECYT grants 1051056, 1061291,
1071125, 3060016, 3070055, 3070057. The generous support to Centro de
Estudios Cient\'{\i}ficos (CECS) by Empresas CMPC is also acknowledged. CECS
is funded in part by grants from the Millennium Science Initiative, Fundaci%
\'{o}n Andes and the Tinker Foundation.


\end{document}